\documentstyle[epsfig]{aipproc}

\input paperdef

\begin{document}

\thispagestyle{empty}
\setcounter{page}{0}
\def\thefootnote{\fnsymbol{footnote}}

\begin{flushright}
BNL--HET--00/47\\
CERN--TH/2000-370\\
hep-ph/0012364 \\
\end{flushright}

\vspace{1cm}

\begin{center}

{\large\sc {\bf Electroweak Precision Tests with GigaZ}}
\footnote{Talk given by S.~Heinemeyer at the 5th International Linear 
Collider Workshop (LCWS 2000), Fermilab, Batavia, Illinois, 24-28 Oct 2000}

%

\vspace{1cm}

{\sc S. Heinemeyer$^{1\,}$%
\footnote{
email: Sven.Heinemeyer@bnl.gov
}%
 and G. Weiglein$^{2\,}$%
\footnote{
email: Georg.Weiglein@cern.ch
}%
}

\vspace*{1cm}

$^1$ HET, Brookhaven Natl.\ Lab., Upton, New York 11973, USA

\vspace*{0.4cm}

$^2$ CERN, TH Division, CH-1211 Geneva 23, Switzerland

\end{center}

\vspace*{1cm}

\begin{abstract}
By running the prospective high-energy \epem\ collider TESLA in the
GigaZ mode on 
the $Z$~resonance, experiments can be performed on the basis of more than 
$10^9~Z$~events. This will allow the measurement of the effective
electroweak mixing angle to an accuracy of 
$\de\sweff \approx \pm 1 \times 10^{-5}$. 
The $W$~boson mass is likewise expected to be measurable with an error of 
$\de\MW \approx \pm 6$~MeV near the $W^+W^-$ threshold. We review the
electroweak precision tests that can be performed with these high
precision measurements within the Standard
Model (SM) and its minimal Supersymmetric extension (MSSM). 
The complementarity of direct measurements at a prospective linear
\epem\ collider and indirect constraints following from 
measurements performed at GigaZ is emphasized. 
\end{abstract}

\def\thefootnote{\arabic{footnote}}
\setcounter{footnote}{0}

\newpage


\title{Electroweak Precision Tests at GigaZ}

\author{Sven Heinemeyer$^*$ and Georg Weiglein$^{\dagger}$}
\address{$^*$HET, Brookhaven Natl.\ Lab., Upton, New York 11973, USA\\
$^{\dagger}$CERN, TH Division, CH-1211 Geneva 23, Switzerland}

\maketitle

\begin{abstract}
By running the prospective high-energy \epem\ collider TESLA in the
GigaZ mode on 
the $Z$~resonance, experiments can be performed on the basis of more than 
$10^9~Z$~events. This will allow the measurement of the effective
electroweak mixing angle to an accuracy of 
$\de\sweff \approx \pm 1 \times 10^{-5}$. 
The $W$~boson mass is likewise expected to be measurable with an error of 
$\de\MW \approx \pm 6$~MeV near the $W^+W^-$ threshold. We review the
electroweak precision tests that can be performed with these high
precision measurements within the Standard
Model (SM) and its minimal Supersymmetric extension (MSSM). 
The complementarity of direct measurements at a prospective linear
\epem\ collider and indirect constraints following from the 
measurements performed at GigaZ is emphasized. 
\end{abstract}

\section{Theoretical basis}
The prospective high-energy \epem\ linear collider TESLA can
be operated on the $Z$~boson resonance
by adding a bypass to the main beam line~\cite{gigazbypass}.
Due to the high luminosity, $\cL = 7 \times 10^{33} \rm{cm}^{-2} \rm{s}^{-1}$, 
about $2 \times 10^9~Z$~events per year can be generated, which will
be referred to as the ``GigaZ'' mode. By using the
Blondel scheme, this results in a measurement of the effective
leptonic mixing angle, $\sweff$, of about 
$\de\sweff \approx \pm 1 \times 10^{-5}$~\cite{moenig}. 
Increasing the collider energy to the $W$-pair threshold,
about \order{10^6} $W$~bosons can be generated resulting in a measurement of 
the $W$~ mass of $\de\MW \approx \pm 6 \mev$~\cite{wwthreshold}.
This increase of precision in $\sweff$ and $\MW$ opens new
opportunities for high precision physics in the electroweak
sector~\cite{gigazsitges,gigaz}.

In this paper we compare the theoretical predictions for $\MW$ and
$\sweff$ in the Standard Model (SM) and the Minimal Supersymmetric
Standard Model (MSSM) with the expected experimental
uncertainties. 
In order to calculate the $W$-boson mass in the SM and the MSSM we use
\BE
\MW^2 = \MZ^2/2 \KKL 1 + \KL 4\pi\al/(\wz \gf \MZ^2)
                           \times (1 + \De r) \KR^{1/2} \KKR~,
\label{mwdeltar}
\EE
where the loop corrections are summarized in $\De r$. 
The quantity $\sweff$ is defined through the
effective couplings $\gvf$ and $\gaf$ of the $Z$ boson to fermions:
\BE
\sweff = 1/(4\,|Q_f|) 
  \KKL 1 - \mbox{Re}\, \gvf / \mbox{Re}\, \gaf \KKR~,
\EE
where the loop corrections 
are contained in $g^f_{V,A}$.
The theoretical input for $\MW$ and $\sweff$ is described in detail in
\citere{gigaz}. It involves corrections up to \order{\al^2}~\cite{sm2lmt4}
and \order{\al\als^2}~\cite{sm3lqcd} in the SM and up to \order{\al\als} in
the MSSM~\cite{mssm2lqcd}.


In the SM the Higgs boson mass is a free parameter. Contrary to this,
in the MSSM the masses of the neutral $\cp$-even Higgs bosons are
calculable in terms of the other MSSM parameters. The largest
corrections arise from the $t$--$\Stop$-sector, where the dominant
contribution reads
$ 
\De \mh^2 \sim \mt^4/\MW^2 
                \log (\mste^2\,\mstz^2/\mt^4) .
$ 
$\mste$ and $\mstz$ denote the two stop mass eigenstates. $\tst$
will later denote the $\Stop$~mixing angle.
Since the \onel\ corrections are known to be very large, we use the
currently most precise \twol\ result based on explicit
Feynman-diagrammatic calculations~\cite{mhiggs2l},
where the numerical evaluation is based on~\citere{feynhiggs}.
The relevant observables together with their uncertainties at various
colliders and their current experimental value can be found in
\refta{tab:precallcoll}. 

\begin{table}[ht!]
\caption[]{\it\footnotesize
Expected precision at various colliders for $\sweff$, $\MW$, $\mt$  and 
the (lightest) Higgs boson mass, $\Mh$. 
``now'' refers to the present accuracy obtained at
LEP, SLD and the Tevatron RunI. ``LHC'' here and in the following
also includes Tev.\ RunII.
See \citere{gigaz} for a detailed list or references.}
\begin{tabular}{|c||c||c|c||c||c||c|}
\cline{2-7} \multicolumn{1}{c||}{}
 & now & LHC & LC  & GigaZ & & 
                                    current central value \\ \hline \hline
$\de\sweff(\times 10^5)$ & 17  & 17  &  6  & 1.3  & & 0.23146    \\ \hline
$\de\MW$ [MeV]           & 37  & 15  & 15  & 6    & & 80.436 GeV \\ \hline
$\de\mt$ [GeV]           & 5.1 & 2   & 0.2 & 0.13 & & 174.3 GeV  \\ \hline
$\de\MH$ [MeV]           & --  & 200 &  50 & 50   & &  -- \\ \hline
\end{tabular}
\label{tab:precallcoll}
\end{table}
\vspace{-1em}


\section{Comparison of SM and MSSM}
\label{sec:SMvMSSM}

\begin{figure}[ht!]
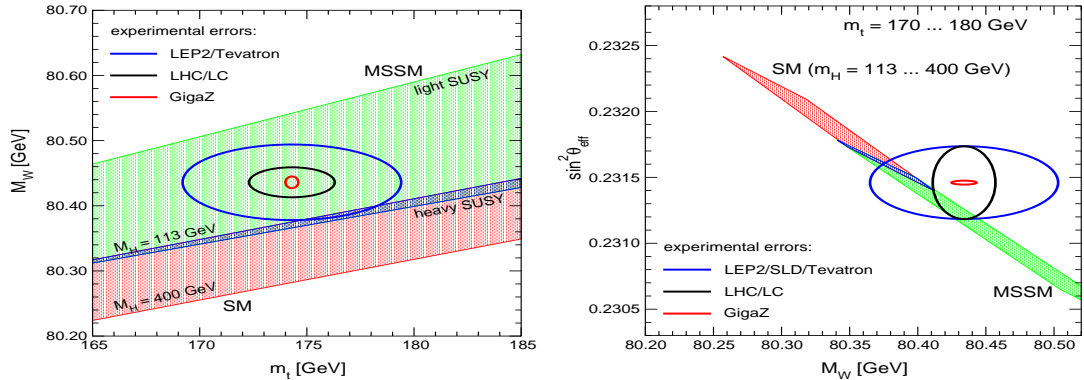

\begin{center}
\mbox{
\epsfig{figure=MWMT01.cl.eps,width=7cm,height=5cm}}
\mbox{
\epsfig{figure=SWMW01.cl.eps,width=7cm,height=5cm}}
\end{center}
\caption[]{
The theoretical prediction of $\MW$ and $\sweff$ is compared to the
experimental measured values with the current LEP/SLD/Tevatron
precision and with the prospective accuracies at the LHC/LC and at GigaZ.
}
\label{fig:SMvMSSM}
\end{figure}

In \reffi{fig:SMvMSSM} the theoretical predictions for $\MW$ and
$\sweff$ obtained in the SM and the MSSM are compared with their
experimental values. 
In the left plot of \reffi{fig:SMvMSSM} the bands in the
$\mt$--$\MW$ plane allowed in the SM and the MSSM are compared to the
(prospective) experimental precisions at LEP/Tevatron, LHC/LC and
GigaZ. The SM band arises from the unknown value of the Higgs boson
mass, where the upper boundary is obtained from the lower bound set by
LEP, $\MH \gsim 113 \gev$~\cite{mhiggsexp}. The band in the MSSM is
due to the unknown masses of the SUSY particles. The upper boundary
corresponds to light SUSY, the lower boundary corresponds to heavy
SUSY, i.e.\ the MSSM is SM like. In the overlap area the SM 
has a Higgs boson in the SUSY range, i.e.\ $\MH \lsim 130 \gev$. 
The plot shows a slight preference of the present data for the MSSM at the
68\%~CL. 

The right plot of \reffi{fig:SMvMSSM} shows the $\MW$--$\sweff$
plane. The allowed area in the SM and the MSSM is compared with the
experimental precision at LEP/SLD/Tevatron, LHC/LC and GigaZ. For the
SM area, the Higgs boson mass has been varied between 
$113 \gev \leq \MH \leq 400 \gev$. The top quark mass has been varied
between $170 \gev \leq \mt \leq 180 \gev$. Both models possess an
allowed parameter space at the 68\%~CL. 

\section{Indirect constraints from GigaZ}

\begin{figure}[ht!]
\begin{center}
\mbox{
\epsfig{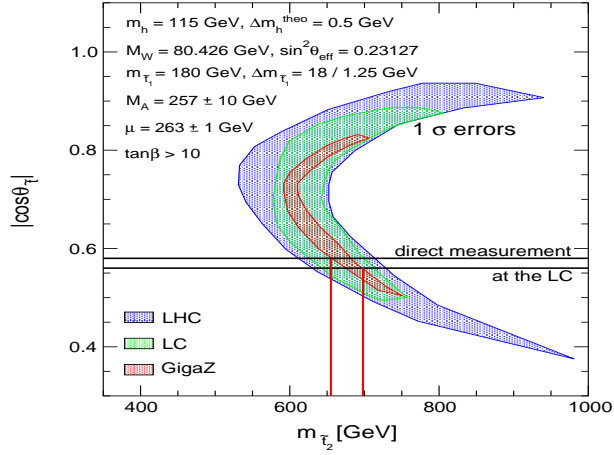}}
%
\end{center}
\caption[]{
Indirect constraints on the MSSM parameter space in the 
$\mstz$--$|\costt|$ plane from measurements of 
$\mh$, $\MW$, $\sweff$, $\mt$ and $\mste$ in view of the 
prospective accuracies for
these observables at a LC with and
without GigaZ option and at the LHC. The direct information on the
mixing angle from a measurement at the LC is indicated together with the
corresponding indirect determination of $\mstz$.
}
\vspace{-2em}
\label{fig:LCvGigaZ}
\end{figure}

Often the indirect constraints on observables obtained at GigaZ
could be complementary to their direct measurements at the Tevatron
RunII, the LHC or at an
LC. As an example we present an analysis for the scalar top sector.
The direct information on the stop sector parameters, $\mste$ and
$\tst$, can be obtained 
from the process $e^+e^- \to \tilde t_1 \tilde t_1$ to a
precision of \order{1\%}~\cite{lcstop}. These direct measurements can be
combined with the indirect information from requiring consistency of the
MSSM with a precise measurement of the 
Higgs boson mass, $m_h$, and the electroweak precision observables. 
This is shown in Fig.~\ref{fig:LCvGigaZ}, where the allowed 
parameter space according to measurements of $\mh$, $\MW$ and $\sweff$
are displayed in the plane of the heavier stop mass, $\mstz$,
and $|\costt|$ for the accuracies at a LC with
and without the GigaZ option and at the LHC (see \refta{tab:precallcoll}). 
For $\mste$ the central value and experimental error
of $\mste = 180 \pm 1.25 \gev$ are taken for LC/GigaZ, while for the LHC an 
uncertainty of 10\% in $\mste$ is assumed. The other parameters
have been chosen according to the mSUGRA reference
scenario~2~\cite{msugrapoints}, with the following 
accuracies: $\MA = 257 \pm 10$~GeV, $\mu = 263 \pm 1$~GeV,
$M_2 = 150 \pm 1$~GeV, $\mgl = 496 \pm 10$~GeV. For the
top-quark mass an error of 0.2~GeV has been used for GigaZ/LC and 
of 2~GeV for the LHC. For $\tb$ a lower bound of $\tan\beta > 10$
has been taken. 
For the future theory uncertainty of $\mh$ from unknown higher-order
corrections an  
error of $0.5$~GeV has been assumed. 
The central values for $\MW$ and $\sweff$ have been chosen in accordance
with a non-zero contribution to the precision observables from SUSY
loops. 

As one can see in Fig.~\ref{fig:LCvGigaZ}, the allowed parameter space in the
$\mstz$--$|\costt|$ plane is significantly reduced
from the LHC to the LC, in particular in the GigaZ scenario. Using 
the direct information on $|\costt|$ from \citere{lcstop}
allows an indirect determination of $\mstz$ with 
a precision of better than 5\% in the GigaZ case. By comparing this indirect 
prediction for $\mstz$ with direct experimental information on
the mass of this particle, the MSSM could be tested at its quantum level
in a sensitive and highly non-trivial way.

\end{document}